\documentclass[aps,prl,twocolumn,superscriptaddress,
nobibnotes,nodoi,noeprint,longbibliography]{revtex4-2}
\usepackage{graphicx}
\usepackage{dcolumn}
\usepackage{bm}
\usepackage{amsmath}
\usepackage{float}
\usepackage{color}

\begin{document}
\title{Hydrodynamics is Needed to Explain Propulsion in Chemophoretic Colloidal Rafts}
\author{Dolachai Boniface}
\thanks{Both authors equally contributed to this work}
\affiliation{Departament de F\'{i}sica de la Mat\`{e}ria Condensada, Universitat de Barcelona, 08028 Spain}
\author{Sergi G. Leyva}
\thanks{Both authors equally contributed to this work}
\affiliation{Departament de F\'{i}sica de la Mat\`{e}ria Condensada, Universitat de Barcelona, 08028 Spain}
\affiliation{Universitat de Barcelona Institute of Complex Systems (UBICS), Universitat de Barcelona, Barcelona, Spain}
\author{Ignacio Pagonabarraga}
\affiliation{Departament de F\'{i}sica de la Mat\`{e}ria Condensada, Universitat de Barcelona, 08028 Spain}
\affiliation{Universitat de Barcelona Institute of Complex Systems (UBICS), Universitat de Barcelona, Barcelona, Spain}
\author{Pietro Tierno}
\email{ptierno@ub.edu}
\affiliation{Departament de F\'{i}sica de la Mat\`{e}ria Condensada, Universitat de Barcelona, 08028 Spain}
\affiliation{Universitat de Barcelona Institute of Complex Systems (UBICS), Universitat de Barcelona, Barcelona, Spain}
\affiliation{Institut de Nanoci\`{e}ncia i Nanotecnologia, Universitat de Barcelona, Barcelona, Spain}
\date{\today}
\begin{abstract}
Active particles driven by 
a chemical reaction are the subject of intense research to date
due to their rich physics, 
being intrinsically far from equilibrium, and their multiple technological applications.
Recent attention in the field is now shifting towards exploring the fascinating dynamics of mixture of active and passive systems. 
Here we realize active colloidal rafts, composed of a single catalytic particle encircled by several shells of passive microspheres
assembled via light activated, chemophoretic flow. 
We show that considering only diffusiophoresis can explain the cluster kinetics but not the cluster  propulsion behavior.
Thus, using the Lorenz reciprocal theorem, we show that propulsion emerges by considering hydrodynamics via the 
diffusioosmotic answer of the substrate to the generated chemophoretic flow. While diffusioosmotic flows are often relegate to a secondary role, our work demonstrates their importance to understand the rich physics of active catalytic systems.
\end{abstract}
\maketitle
\textbf{\textit{Introduction.-}}
In the past few years, active colloidal particles have led to several exciting developments in the field of non-equilibrium statistical mechanics~\cite{Ramaswamy2010,Marchetti2013,Fodor2016,Nardini2017} while being also used as simplified models to reproduce emerging phenomena in biological self-propelling systems~\cite{Elgeti2015,Bechinger2016,Zottl2016,Gompper2020}. Since the pioneering works of Ismagilov \textit{et al.}~\cite{Ismagilov2002} and Paxton \textit{et al.}~\cite{Pax04}, 
chemical reactions have been routinely used 
to induce propulsion in asymmetric systems~\cite{Wang2006}
including 
Janus particles~\cite{Jonathan2007,Das2015,Theurkauff2012,Ketzetzi2020}, nanorods~\cite{Yang2006,Wang2013}, dimers~\cite{Gunnar2007,Valadares2010}, mixtures~\cite{Niu2017,Agudo-Canalejo2019} and many others~\cite{Soto2014,Pohl2014,Jang2016}.
Besides the interest in the reaction mechanism that leads to net motion, 
these particles 
showed the capabilities to pick up,  transport,  and release microscopic  cargoes~\cite{Jared2008,Baraban2012,Palacci2013,Jang2021}. Thus, they 
may find direct applications 
in different technological fields, including biomedicine~\cite{Nelson2010},  targeted  drug  delivery~\cite{Kim2013}  and  microfluidics~\cite{Sanchez2011}.
 
In most of these catalytic systems, self-propulsion is usually explained in terms of diffusiophoresis or chemophoresis, namely the particle motion in a concentration gradient~\cite{Anderson1989}. However, in presence of a gradient  
also near a fixed surface, such as the substrate near a particle, there will be an osmotic 
flow~\cite{Marbach2019}. For active systems near a substrate, this osmotic flow may affect the system dynamics through viscous interactions~\cite{Palacci20132}. Indeed, the osmotic flows on the substrate may compete with 
particle diffusiophoresis.
Because both phenomena have a similar osmotic origin, the diffusiophoresis and substrate diffusioosmosis contributions are difficult to disentangle \cite{Liebchen2021}. 
Thus, most of the theoretical
and simulation models in the field do not consider the
impact of hydrodynamic interactions associated to diffusioosmosis.
In contrast, a recent theoretical work showed
that the diffusioomotic contribution  
in active Janus particles
can be even used to guided a interaction with a chemically
patterned substrate~\cite{Uspal2016}.



Here, we combine experiments and theory to demonstrate that the diffusioosmotic flow induced by the catalytic particle due to the near surface is necessary to describe the motion of active particles driven by chemical reactions.
We realize active colloidal rafts composed of several shells of passive spheres around a single catalytic apolar particle.
These clusters grow up to 
an area of $80$ times the 
active inclusion, corresponding to
$7$ compact shells of passive spheres, and investigate the raft kinetics and dynamics during the illumination process. 
We find that the clusters display self-propulsion despite being made of
symmetric shells of passive spheres.
Numerical simulations based only on a 
purely diffusiophoretic
system, without osmotic flow on the substrate, 
reproduce
the raft kinetics but not the cluster direction of
motion and its persistence length.
We show that 
hydrodynamics and the close boundary are 
essential features that should be taken into account 
to explain the mechanism of motion of the composite clusters. 

\begin{figure}[h]
\begin{center}
\includegraphics[width=\columnwidth,keepaspectratio]{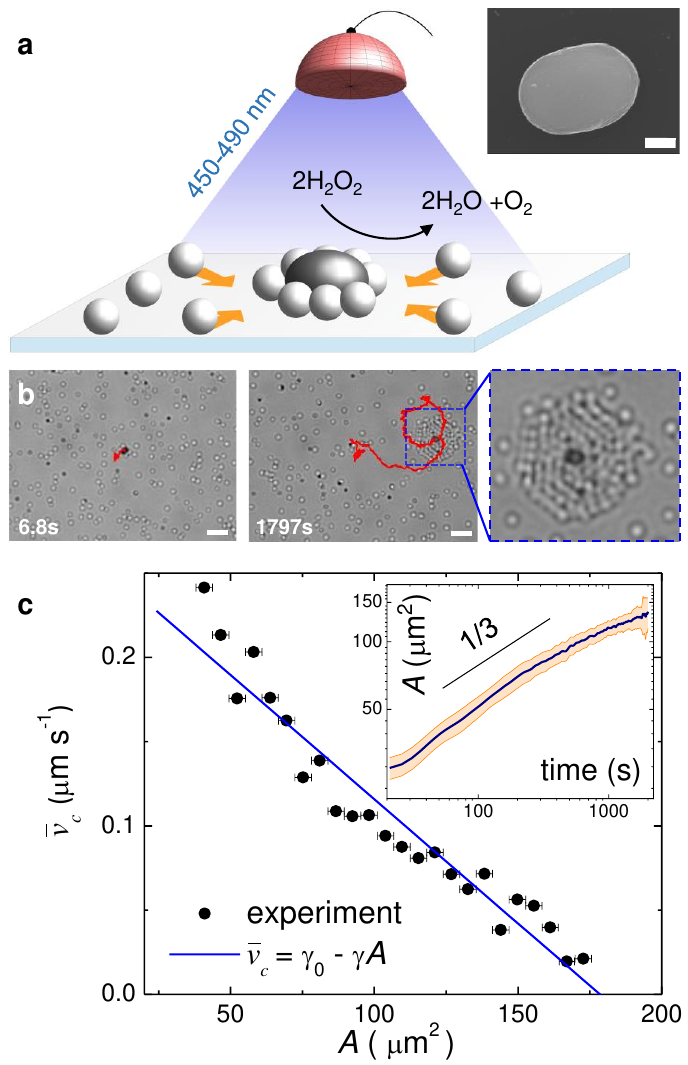}
\caption{(a) Scheme showing the assembly of the colloidal raft.  
Top inset shows an electron microscopy image of one hematite, scale bar is $500$ nm. 
(b) Sequence of two images of a growing raft with superimposed (red) the trajectory of the central active particle. Time $t=0$s corresponds to light application. Scale bar is $5 \, \rm{\mu m}$. Last image displays the final cluster size, see VideoS1 in~\cite{EPAPS}.
(c) Average raft velocity $\bar{v}_{c}$ versus cluster area $A$ 
showing the experimental data (black disk) and a linear regression with $\gamma_0= 0.26\pm 0.02\, \rm{\mu ms^{-1}}$ and a negative slope $\gamma=(1.48 \pm 0.02)\cdot 10^{-3}\rm{\mu m^{-1}s^{-1}}$. Inset shows a log-log plot of the area versus time 
for several rafts, error bars are indicated by the shaded red region.}
\label{figure1}
\end{center}
\end{figure}

\textbf{\textit{Experiments.-}} 
Our colloidal rafts are realized by 
illuminating with blue light (wavelength $\lambda =  450-490$nm)
synthesized hematite ellipsoids with short and long axis equal to $1.3 \, \rm{\mu m}$ and $1.8 \, \rm{\mu m}$ resp, inset Fig.~\ref{figure1}(a). These particles are dispersed with passive silica spheres (1 $\mu$m diameter) in an aqueous solution of hydrogen peroxide (3.6 \% w/v).
The pH solution is raised to $\sim 9.2$ by adding Trimethylphenylammonium to make the hematite hydrophilic due to hydroxylation of its surface~\cite{Shrimali2016}.
The colloidal dispersion is sediment over a
glass substrate of a sealed rectangular capillary tube. The relative density is below $1$ active particle for $2000$ passive ones, with a total surface fraction of $\sim 6\%$. Once the light is applied, 
the hematite particles start the decomposition of hydrogen peroxide in water, following the reaction: $2$H$_2$O$_{2(l)}\rightarrow$ O$_{2(g)}$+2H$_2$O$_{(l)}$, Fig.~\ref{figure1}(b). It was previously shown that a such chemical reaction induced propulsion in 
Janus colloids with anisotropic coating~\cite{Ebbens2018,Popescu2018}.
For a single hematite particle we find that diffusiophoresis induces an enhanced diffusive dynamics as shown in the Supplementary Material (SM) in~\cite{EPAPS}.
The presence of a near passive sphere induces a strong phoretic attraction which 
generates a stable and large passive-active cluster displaying self-propulsion~\cite{Codina2022}.
We find that the rafts follow a sub-linear growth with a power law behavior up to $t=2000$ s ($\simeq$ 0.6 hours), inset in Fig.~\ref{figure1}(c). 
The exponent $1/3$ is consistent with the Ostwald coarsening process, as described by the Lifshitz-Slyozov-Wagner theory~\cite{Bray1994}. Such exponent was predicted in scalar field theory of active systems~\cite{Wittkowski2014} and recently experimental observed in clustering passive particles by active agents~\cite{Bouvard2023}.
During growth the raft  translates and rotates, and the  association of both can result in looping trajectories, Fig.~\ref{figure1}(b). The system accumulates up to $6-7$ layers of passive particles for one-hour experiment.
The mean cluster velocity, $\bar{v}_{c}$ linearly decreases with the cluster area $A$, 
reducing almost to zero for the largest size of $A=175 \, \rm{\mu m^2}$, Fig.~\ref{figure1}(c). 

\begin{figure}[b]
\includegraphics[width=\columnwidth]{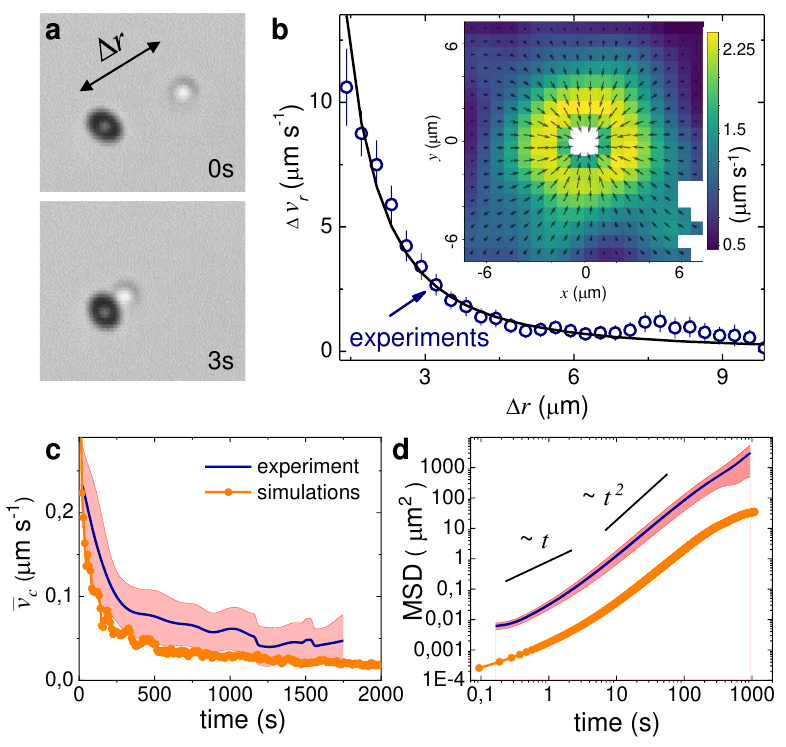}
\caption{(a) Sequence of images showing the attraction of a silica particle towards the hematite 
once blue light is applied ($t=0)$.
(b) Relative speed $\Delta v_r$ versus relative distance $\Delta r$: the solid line is fitted to the data (blue circles) following Eq.~\ref{relative_speed}. Inset displays a heat map of the velocity and direction of a passive particle near the hematite. (c,d) Mean cluster speed $\bar{v}_{c}$ (c) and mean square displacement (MSD) (d) versus time
from experiments (blue line) and simulation (orange disks).  In both graphs the shaded red regions denote experimental uncertainties.}
\label{figure2}
\end{figure}

\textbf{\textit{Simulations.-}} To understand the kinetics and self-propulsion behavior, we first perform Brownian dynamic simulations
using input parameters obtained from the experimental data. Here we assume a purely diffusiophoretic system.
We consider a bath of $i=1..N$ passive particles at positions $\bm{R}_i$ (diameter $\sigma_p$, surface mobility $\mu_p$ and diffusion coefficient $D_p$) with an unique active particle. 
To model the aspect ratio of the experimental ellipsoids, the hematite
is considered as a dumbbell of two active particles, $\alpha={1,2}$, at positions $\bm{r_{\alpha}}$ (diameter $\sigma_a=1.3 \rm{\mu m}$, surface mobility $\mu_a$, and diffusion coefficient $D_a$) joined by a spring with rest length 0.5 $\mu$m, and  force of magnitude $F^h$ along the vector $\hat{\bm{n}}_i=({\bf r}_i-{\bf r}_j)/r_{ij}$ joining the two beads.  Thus,  we integrate the overdamped Langevin equations:
\begin{eqnarray}
\dot{\bm{r}}_{\alpha}&=&\bm{v}_{\alpha}+(F^h\hat{\bm{n}}_{\alpha}+\bm{F}^{c}_{\alpha})/\gamma_a+\sqrt{2D_a}\bm{\xi}_{\alpha} \, \, \, ,\\
      \dot{\bm{R}}_i &=& \bm{V}_i + \bm{F}_i^c/\gamma_p+\sqrt{2D_p}\bm{\xi}_i\, \, \, .
\label{simu}
\end{eqnarray}
where $\gamma_a$ and $\gamma_p$ correspond to the active and passive friction coefficients, respectively. Here $\bm{F}_i^c$ and $\bm{F}_{\alpha}^c$ account for steric forces given by a Weeks-Chandler-Andersen potential, which prevent passive and active particles from overlapping. The term $\bm{\xi}_i$ is a random Gaussian noise that accounts for the thermal bath.  
Each bead constituting the dumbbell in the hematite acts as a source ~\cite{Golestanian2007,Soto2014,Codina2022} of a chemical field, $\phi$. A  second particle with  mobility $\mu_p$ ($\mu_a$) will experience a slip velocity on its surface,  $\bm u_s = \mu_p(\mu_a) \nabla_\parallel \phi$, that leads to a net diffusiophoretic velocity
 $\bm{V}_i$ ($\bm{v}_{\alpha}$), see~\cite{EPAPS} for the derivation. Accordingly,  
the relative speed of approach $\Delta v_r$ between an active and a passive particle at a relative distance $\Delta r$ reads,
\begin{equation}
\Delta v_r=v_{\alpha} + V= 
v_0 \left[ \bar{\mu }   \left( \frac{\sigma_a}{\Delta r}\right)^2 +\frac{1}{4} \left( \frac{\sigma_p}{\sigma_a} \right)^3 \left( \frac{\sigma_a}{\Delta r} \right)^5 \right]   \, \, \, ,
\label{relative_speed}
\end{equation}
where $\bar{\mu} = \mu_p/\mu_a$ is the ratio of the two mobilities.
The detailed derivation of this functional form is provided in~\cite{EPAPS}. We use Eq.~\ref{relative_speed} to fit the 
experimental data as shown in Fig.~\ref{figure2}(b),
and extract a characteristic diffusiophoretic velocity given by $v_0=11.6\pm 0.4$ $\mu$m $s^{-1}$.
Note that the heat map of the velocity field shown in the inset in Fig.~\ref{figure2}(b) becomes slightly anisotropic (less than $5\%$) if the orientation of the hematite is kept fixed with a constant field, as shown in~\cite{EPAPS}.
We also note that the attraction 
between the passive and active particle is only possible if
$\mu_p$ is negative.
More details on the other terms used in Eq.~\ref{simu} and on the simulations are given in~\cite{EPAPS}.

\begin{figure}[t]
\includegraphics[width=0.95\columnwidth]{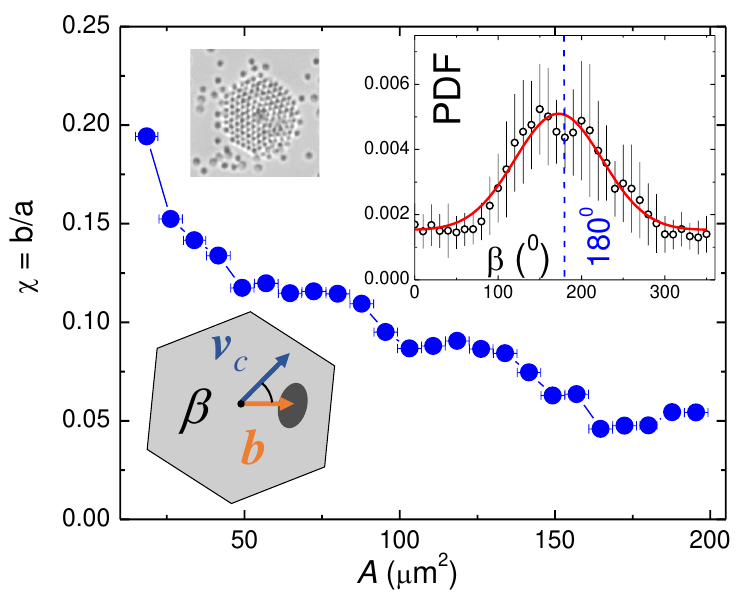}
\caption{Experimentally measured asymmetry parameter $\chi=b/a$ versus cluster area $A$, being $a$ the cluster radius. 
Top left inset shows an image of a cluster, while right inset displays the distribution of angles $\beta$ between 
the cluster velocity $v_{c}$ and the vector $\bm{b}$ pointing from the cluster center to the hematite particle.
These quantities are defined in the schematic in the bottom inset.}
\label{figure3}
\end{figure}

The simulations explain 
some of the experimental features: the growth 
of the raft area as $t^{1/3}$, 
the emergence of the self-propulsion behavior and, 
in particular, the decrease 
of the raft velocity with the cluster area as shown in Fig.~\ref{figure2}(c).
However, when comparing the raft dynamics via other observables, we find already some discrepancies. For example in Fig.~\ref{figure2}(d) we show the average translational mean square displacement MSD$(\tau) \equiv \langle (\bm{r}(t)-\bm{r}(t+\tau))^2 \rangle \sim \tau^{\delta}$, with $\tau$ the lag time and $\langle … \rangle$ a time average. Via the exponent $\delta$, the MSD can be used to distinguish the diffusive ($\delta = 1$) dynamics from sub-[super] diffusive ($\delta < 1$ [$\delta > 1$]) and ballistic one ($\delta = 2$).
We define the persistence length of the trajectory $l_p$, as the characteristic length over which the velocity orientation decorrelates. We calculate this quantity from the cluster trajectory as, $\langle \cos (\theta_v(d+\Delta l)-\theta_v(d) \rangle_d \propto\exp (-\Delta l/l_p)$ being $d$  the distance travelled by the cluster and $\theta_v$ the  orientation of the velocity vector. From the experiment, we measure a persistence length $l_p\simeq 20 \, \rm{\mu m}$ which is significantly larger than the one predicted in the simulations,  $l_p\simeq 2.5 \, \rm{\mu m}$. As we show below, 
this discrepancy can be solved by considering 
the asymmetric location of the hematite within the cluster.

\textbf{\textit{Cluster asymmetry.-}} 
To better understand the origin of the raft propulsion, we have analyzed in detail the position of the hematite source within the cluster.  During the growing process and in the steady state we find that the hematite is not exactly located in the 
cluster' geometric center, but it is displaced a small distance $b$. As shown in Fig.~\ref{figure3}, the asymmetry parameter $\chi=b/a$, being $a$ the radius of the cluster, decreases with the raft area $A$. Moreover, the analysis of the 
distribution of angle $\beta$ between the cluster velocity $v_{c}$
and the asymmetry vector $\bm{b}$ gives further insight on the propulsion direction.
As shown in the inset of Fig.~\ref{figure3},
such wrapped distribution is Gaussian (red line) and centered around $\beta = 180^{\circ}$, meaning that the raft propels with the active particle at the rear. Numerical simulations show that the clusters instead tend to propel 
with the active particle at the front, as shown in the SM, VideoS2. 

Qualitatively, we can understand how
the asymmetric location of the hematite in the cluster
impacts the persistence length. When a colloidal
raft moves in a crowded environment of passive particles, they tend to accumulate at the front. Thus, a cluster moving with the
hematite shifted toward the front has to change regularly its motion direction to maintain this configuration, as reported in the simulations. 
While for a cluster moving with the hematite
shifted towards the rear, the colloids
front accumulation preserves the asymmetry and the motion direction, as observed in the experiments. The two situations lead respectively to
a system with a relatively low and high persistence length. To confirm this hypothesis, we have implemented a specific simulation by imposing that the cluster moves with the hematite
at the rear. As shown in VideoS3 in \cite{EPAPS}, we observe
a much longer persistence length, closer to the experimental results. 

The discrepancy between the numerical and experimental results arises from the assumption that the system is purely diffusiophoretic. The simulation neglects hydrodynamics and does not consider the presence of the near wall. Indeed in  a separate set of experiments we have replaced the glass substrate with a polystyrene one and have observed a decrease of the cluster area, as shown in~\cite{EPAPS}. This effect highlights the importance of the bottom surface.

\begin{figure}[t]
\includegraphics[width=\columnwidth]{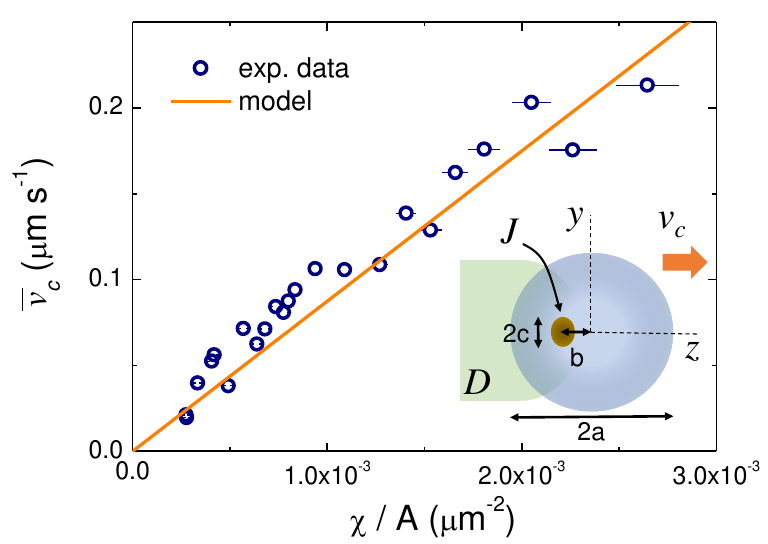}
\caption{Experimental data of the mean cluster velocity $\bar{v}_{c}$ versus ratio $\chi/A$ being $\chi=b/a$.
Scattered circles are experimental data while the continuous line is a fit from the model, see Eq.~\ref{eq:conc_osmotic_model} in the text. 
Inset illustrates a schematic of the model: the cluster is considered as a  thin disk of radius $a$ with an active source of size $\sigma_a$ and distance $b$ from the center. $J$ and $D$ denote respectively the release rate of the source and solvent diffusion rate.}
\label{figure4}
\end{figure}

\textbf{\textit{Theory.}} To include the effect of 
hydrodynamics and the proximity of the wall, 
we approximate the colloidal raft by a disk 
of diameter $2a$ and the shifted hematite by a "semi-punctual'' source, where the concentration field $\phi$ is similar to a punctual source except along the source surface, where $\phi$ is constant.   
We orient the system such that the unit vector 
$\bm{e}_z$ is diametrically opposed to the vector $\bm{b}$
linking the cluster center to the source. The negative or positive sign of the cluster velocity $v_{c}$ indicates a disk moving with the source at the front or the rear, respectively.
We assume that the catalyzed product is released at the rate $J$, and diffuses in bulk with a diffusion coefficient $D_c$. We consider two parallel surfaces, the disk ($p$) and the substrate ($S$), separated by $h$, such as $h/a\ll1$.
To describe the disk dynamic we introduce two dimensionless numbers: the P\'eclet  $\text{Pe}_c=\frac{v_{c}a}{D_c}$, the Damk\"ohler number $\text{Da}=\frac{\mu_{p} J}{4\pi a D_c^2}$
which relates the reaction rate to the diffusive mass transport rate.
Experimentally, $\text{Pe}_c\simeq 10^{-4}\ll1$ thus the transport of the solute is dominated by diffusion, and the source motion can be disregarded. Therefore at a distance $r$ from the source the chemical gradient is $\nabla \phi = -J/(4\pi D_c r^2)\bm{e}_r$.
 The concentration gradient generates a slip osmotic flow $ \bm u_S=\mu \nabla_S \phi, $ along the relevant surfaces, namely  the disk surface $p$ and the substrate $S$, such that
$\left. \bm u\right|_p=v_{c} \bm e_z+\mu_p \bm \nabla \phi$, and $\left. \bm u\right|_S=\mu_S \nabla \phi$.
The disk motion is force-free, hence
$\bm{F}_v+\bm{F}_{p}+\bm{F}_S=0$, where $\bm{F}_v$ is the damping force due to the motion of the disk, 
$\bm{F}_{p}$ is the phoretic force associated with the slip velocity on the disk's surface, and 
$\bm{F}_S$ the osmotic contribution coming from the slip velocity on the wall.
See~\cite{EPAPS} for details of all terms employed and the extended model. 

Using the Lorentz reciprocal theorem, we  arrive at
\begin{equation}
\text{Pe}_c\simeq 2 \text{Da} (1-\mu_S/\mu_p) \chi + O(\chi^2) \, \, \, ,
\end{equation}
and, accordingly, the  velocity of the disk at the first order in $\chi$ is given by
\begin{equation}
v_{c}\propto (\mu_{p}-\mu_S)\frac{\chi}{A}. \label{eq:conc_osmotic_model}
\end{equation}
Note that if we remove the osmotic flow along the substrate, the term $\mu_S$ disappears from Eq.~\ref{eq:conc_osmotic_model}, and $
v_{c}\propto \mu_{p}\frac{ \chi}{{A}}$. 
Neglecting or taking into account this flow leads almost to the same dependencies with $\frac{ \chi}{{A}}$ for the velocity of the disk which is consistent with the experimental observation, Fig.~\ref{figure4}. The difference between the osmotic mobilities $\mu_{p}-\mu_S$ in Eq.
\ref{eq:conc_osmotic_model} marks of the competition between diffusiophoresis and substrate diffusioosmosis. It controls the sign of $v_c$, i.e. direction of motion of the raft.
Since the passive colloid and the substrate are made of silica, it is reasonable to assume that $\mu_S$ is comparable to $\mu_p$. We also deduce from the clustering phenomenon that $\mu_p<0$.
If we assume that ${\mu}_S/\mu_p>1$, the osmotic model in Eq.~\ref{eq:conc_osmotic_model} predicts a cluster moving with the hematite at the rear, as we observe experimentally. 

\textbf{\textit{Conclusion.-}}
We have investigated the dynamics of
active colloidal rafts composed of a central hematite particles
and several shells of passive colloids.
We have shown that this system displays a clustering phenomenon due to diffusiophoresis, and collective self-propulsion resulting from  diffusioosmosis on the nearby substrate. Indeed, simulations based only on diffusiophoresis describe well the clustering kinetics, but cannot explain the cluster direction of motion and persistence length. Our model solves the discrepancy by considering the cluster asymmetry and, in particular, the substrate diffusioosmotic flow. Thus, we have shown that there is a competition between the diffusiophoresis and osmosis for the cluster motion, and the crucial role of the substrate diffusioosmotic flow on the dynamics. In the line of these results, previous works in the field have also shown the importance of considering the osmotic flow generated by an active particle close to a wall~\cite{Das2015,Katuri2021}. The theoretical approach based on the Lorentz reciprocal theorem, could be extended to many other catalytic active systems close to a substrate, taking into account the proper boundary conditions. In our experiments, we approximate the raft to a disk allowing to reach an analytical expression that captures the underlying physics of this complex, yet rich hybrid active passive system.

This work has received funding from the European Research Council (ERC) under the European Union’s Horizon 2020 Research and Innovation Programme (grant agreement no. 811234). S.G.L. and I.P. acknowledge support from Ministerio de Ciencia, Innovaci\'on y Universidades (grant no. PID2021-126570NB-100 AEI/FEDER-EU) and from Generalitat de Catalunya under project 2021SGR-673. P.T. and I.P. acknowledge support from the Generalitat de Catalunya (ICREA Acad\'emia).

\end{document}